**Malaria elimination campaigns in the Lake Kariba region of Zambia: a spatial dynamical model**

Milen Nikolov PhD[1], Caitlin A. Bever PhD[1], Alexander Upfill-Brown MSc[1], Busiku Hamainza PhD[2], John M. Miller PhD[3], Philip A. Eckhoff PhD[1], Edward A. Wenger PhD[1], Jaline Gerardin PhD[1]*

[1]Institute for Disease Modeling, Bellevue, WA, United States
[2]National Malaria Control Centre, Lusaka, Zambia
[3]PATH Malaria Control and Elimination Program in Africa (MACEPA), Lusaka, Zambia

*To whom correspondence should be addressed.
Email: jgerardin@intven.com
Telephone: +1 425 691 3343
3150 139th Ave SE
Bellevue WA 98005
USA

**Abstract**

Background

As more regions approach malaria elimination, understanding how different interventions interact to reduce transmission becomes critical. The Lake Kariba area of Southern Province, Zambia, is part of a multi-country elimination effort and presents a particular challenge as it is an interconnected region of variable transmission intensities.

Methods

In 2012-13, six rounds of mass-screen-and-treat drug campaigns were carried out in the Lake Kariba region. A spatial dynamical model of malaria transmission in the Lake Kariba area, with transmission and climate modeled at the village scale, was calibrated to the 2012-13 prevalence survey data, with case management rates, insecticide-treated net usage, and drug campaign coverage informed by surveillance. The model was used to simulate the effect of various interventions implemented in 2014-22 on reducing regional transmission, achieving elimination by 2022, and maintaining elimination through 2028.

Findings

The model captured the spatio-temporal trends of decline and rebound in malaria prevalence in 2012-13 at the village scale. Simulations predicted that elimination required repeated mass drug administrations coupled with simultaneous increase in net usage. Drug campaigns targeted only at high-burden areas were as successful as campaigns covering the entire region.

Interpretation

Elimination in the Lake Kariba region is possible through coordinating mass drug campaigns with high-coverage vector control. Targeting regional hotspots is a viable alternative to global campaigns when human migration within an interconnected area is responsible for maintaining transmission in low-burden areas.

Funding

Bill and Melinda Gates through the Global Good Fund and the Bill and Melinda Gates Foundation.

**Research in context**

Evidence before this study

As malaria transmission declines in many regions around the world, more countries are developing plans for elimination. In the southern Africa region, elimination within the next decade will likely require a combination of vector control and mass drug campaigns. Mathematical models have been developed to guide policy and describe features of successful campaigns. A PubMed search on March 7, 2016 of ((malaria[Title]) AND (model[Title] OR modeling[Title] OR modelling[Title]) AND elimination) for all articles to date yielded 52 results, of which 14 were primary research articles on dynamical malaria models used to investigate elimination. Of the 14, five modeled a specific geographical region, two were spatial models at the province level, and one included human mobility. All models predict mass drug administration to be more effective at reducing transmission than mass screen-and-treat.

Added value of this study
This is the most comprehensive effort to date to capture and model malaria transmission within any specific geographical context, and by modeling at the village level, this is also the finest-resolution spatial model of malaria. For the first time, this study addresses the requirements for elimination in a non-generalized setting whose interconnectedness and spatial variation in transmission present substantial barriers to achieving elimination.

Implications of all the available evidence
Elimination is possible in this challenging region, and thus likely throughout southern Africa, if mass drug administrations are coupled with simultaneous high-coverage vector control. Targeting regional hotspots of transmission is a viable way to conserve resources without sacrificing campaign success.

**Introduction**

Malaria is a vector-borne parasitic disease affecting millions of people worldwide, with *Plasmodium falciparum* still causing over 400,000 deaths per year.[1] Recent escalation in vector control has greatly reduced global burden and brought many regions close to elimination.[2] In some settings, mass drug campaigns have been an effective tool for depleting the human infectious reservoir and breaking the cycle of transmission, although the effectiveness of such campaigns has been mixed.[3]

The ultimate goal is global malaria eradication, but the extreme heterogeneity in transmission levels, vector bionomics, health systems, and population densities limit the applicability of any single strategy.[4] Elimination at a single country level could be challenging to maintain in the presence of high cross-border movements of infected individuals, depending on the country and its regional context.[4] As a result, the concept and implementation of regional malaria eliminations provides a useful staging for progress towards eventual global eradication.[5]

Southern Africa is one region where programs are planning operational strategies for elimination.[6] Elimination in southern Africa requires elimination in the Lake Kariba region of Southern Province, Zambia, where areas of high- and low-intensity transmission are interconnected. Understanding how to achieve elimination in the microcosm of the Lake Kariba area would provide a solution to how to achieve elimination in Southern Africa and possibly in a number of other challenging settings.

The Zambia National Malaria Control Centre has successfully scaled up recommended malaria control interventions over the past decade and is pursuing alternative methods to further reduce malaria transmission, including community-targeted parasite reservoir reduction strategies.[7,8] Beginning in 2012, mass drug campaigns have been carried out annually in the Lake Kariba region of Southern Province, Zambia (Figure 1), where transmission is seasonal and spatially variable.

Understanding the small-scale variation in the interconnected Lake Kariba region through mathematical modeling provides important insights into the critical thresholds for successful outcomes. Previous modeling efforts have provided guidance for conducting successful campaigns in generic settings,[9–11] yet limited work has been done on understanding how a specific geography's individual features can also affect campaign outcomes[12,13] or how spatial variation in vectorial capacity can sustain transmission in interconnected areas. In this work, the most detailed spatial model of a specific geography to date is constructed and used to predict how various factors such as variation in transmission intensity and human migration patterns interact to influence the success of control and elimination efforts.

Local transmission dynamics were reconstructed within a mechanistic model of malaria transmission using the high-resolution surveillance data collected during the mass drug campaigns of 2012-13 in the Lake Kariba region, which included infection status, age, GPS coordinates of households, recent symptoms, recent treatment, and insecticide-treated net (ITN) usage. Village-scale biting rates were selected to calibrate local malaria prevalence and incidence to longitudinal surveillance data and seasonal patterns of clinical case counts reported at health facilities respectively.

The simulation framework was then used to assess the outcome of a variety of post-2016 intervention scenarios. Simulations predict that on their own, mass drug administrations (MDAs), high-coverage ITN distributions, and ramp-ups in case management do not result in regional elimination within the next decade. However, elimination is simulated to be likely if these interventions can be deployed simultaneously at high coverage.

**Methods**

*Study site overview*

During each of the 2012 and 2013 dry seasons (June-November) in Southern Province, Zambia, three large-scale mass-screen-and-treat (MSAT) rounds were undertaken. Individuals were visited at their homes in a full community census and, following consent, administered rapid diagnostic tests (RDT). Test-positive individuals were treated with the antimalarial drug artemether-lumefantrine (AL), of which the first dose was directly observed. During each MSAT round, RDT results were geo-tagged by household location. Information was collected on household demographics, ITN usage, recent fevers, and recent drug treatments.

In this analysis, focus was restricted to a contiguous block of twelve health-facility catchment areas (HFCAs) in Gwembe and Sinazongwe Districts along Lake Kariba (Figure 1). These HFCAs cover approximately 80,000 people living in a geographic area spanning a range of endemic malaria transmission intensities.

*Constructing a detailed spatial model of malaria transmission in the Lake Kariba region*

High-resolution spatial simulations were conducted with EMOD v2·0, a mechanistic, individual-based model of malaria transmission including vector life cycle dynamics and within-host immunity.[14,15] Demographics, migration, ITN usage, treatment-seeking, and drug campaign coverage in simulations were informed by survey data collected during the MSAT rounds (Figures S1-S7). Spatial variation in transmission intensity was captured by selecting local vector larval habitat availabilities to match observed seasonal and spatial patterns in RDT prevalence and clinical incidence. Details are provided in Supplemental Methods.

Household locations within the twelve HFCAs were aggregated to 115 village-sized clusters (Figure 1). Migration rates between clusters were approximated by a gravity model based on local roads, estimated travel times, and fraction of longitudinally-linked individuals found in different clusters in subsequent surveillance rounds. Cluster-specific daily temperature, humidity, and rainfall were estimated from weather station data and the NOAA African Rainfall Estimation Algorithm.[16]

ITN usage was determined by cluster-level reported usage in the 2012-13 survey and historical estimates from the Malaria Atlas Project.[2,17] ITN usage among adolescents was modeled at half that of other individuals. Community-level ITN effects were similar between surveillance and simulation data. Case management with AL was modeled uniformly across all clusters beginning in January 2007 according to the mean fraction of individuals with fever within the last two weeks who sought treatment as reported in surveillance data.

Clusters were assigned to one of three drug campaign coverage levels based on average HFCA coverage estimated from longitudinally-linked individuals found during the 2012-13 campaigns. Each MSAT campaign included three rounds with independent coverage and assumed full compliance with drug regimens.

Given cluster-specific ITN usage and MSAT coverage, cluster-specific transmission intensity was determined by sampling over a range of larval habitat availabilities for *Anopheles arabiensis* and *An. funestus*. Both cluster-level prevalence by RDT from the six MSAT rounds in 2012-13 and the December 2011 pilot round as well as weekly clinical case counts from the Zambia National Malaria Control Centre's Malaria Rapid Report system were used to inform the selection of habitat availability at each cluster. Fit quality was evaluated based on the weighted sum of mean square error of prevalence and rank correlation of clinical incidence.

*Predicting outcomes of future interventions*

Simulations up to and including the 2012-13 MSAT activities were extended to model the 2014-15 MDAs carried out in the study area, where simulated MDA campaigns with dihydroartemisinin-piperaquine (DP), each consisting of two rounds separated by 60 days, were administered to all clusters in December 2014 and July 2015. During MDA, individuals received treatment regardless of their RDT result. Simulations were followed until 2028.

Under case management ramp-up, treatment rates for symptomatic malaria increased after 2015 to >90% by 2019. Under ramped-up ITN usage, ITN distributions continued between 2015 and 2022 at historical rates. Under aggressive ITN usage, ITNs were distributed at 80% coverage every two years between 2016 and 2022. All simulated individuals with an ITN used their net every night.

For simulations where MDAs were extended through 2020, two rounds of MDAs with DP were administered beginning each July for the five years from 2016-20. Campaign coverage at each cluster was set at the cluster's 2012-13 MSAT coverage. For targeted MDAs, clusters in high-burden HFCAs received MDAs from 2016-20 while clusters in low-burden HFCAs did not receive MDA after 2015. See Supplemental Methods for details on intervention scenarios.

*Role of the funding source*

The funders had no role in study design, data analysis, decision to publish, or preparation of the manuscript. The corresponding author had full access to all data in the study and had final responsibility for the decision to submit for publication.

## Results

*Malaria transmission in the Lake Kariba region*

A high-resolution spatial model of the Lake Kariba study site was configured based on village-scale clusters of households and ITN usage, MSAT coverage, case management, and human migration rates derived from the 2012-13 RDT prevalence surveys (Figure 1, Figures S1-S7 and Supplemental Methods). Malaria transmission was modeled within each cluster, where vector populations were driven by cluster-specific climate data and local abundance of larval habitats.

Preliminary entomological data indicated that both *Anopheles arabiensis* and *Anopheles funestus* are present in the study area, with *arabiensis* biting rates highest between January and April during the warm rainy season while *funestus* peaks in September at the beginning of the hot dry season (Figure 2A).[18] Relative abundances of *arabiensis* and *funestus* govern the seasonality of malaria transmission, while absolute abundances determine the intensity. For each village-scale cluster, combinations of larval habitat availabilities for *arabiensis* and *funestus* were simulated with appropriate patterns of ITN usage, case management rates, and MSAT coverage. The resulting simulated prevalence and clinical case counts were compared with surveillance data to select the combination of habitats yielding the best fit to field observations (Figure 2B, Supplemental Methods).

Calibration to surveillance data resulted in the expected gradient of higher habitat availability of both vector species closer to Lake Kariba and lower availability in the higher-altitude HFCAs more distant from the lake (Figure 2C). Chiyabi HFCA was predicted to have the largest amount of *funestus* in this region, suggesting that ITNs may be particularly effective there as *funestus* is indoor-feeding and highly anthropophilic.[19]

The calibrated simulations captured both fine-scale cluster-level variation in malaria prevalence and large-scale spatio-temporal trends of temporary reduction in prevalence observed in the study area following MSAT rounds (Figure 3A, Figure S8). Surveillance in 2012-13 reported stalled reduction in prevalence by late 2012 and rapid rebound after the following rainy season. This observation was replicated in the spatial model, where significant re-infection between rounds two and three in lakeside clusters drives rebound throughout the study area.

The observed seasonality of weekly RDT-confirmed clinical cases was well-captured by simulations, including a characteristic pattern of high case counts between December and June and a small rise in cases in Chipepo and Sinamalima HFCAs as temperatures rise in October (Figure 3B). However, in many HFCAs simulations consistently underestimated the magnitude of observed case counts. While simulations estimate a constant proportion of clinical cases seek care at health facilities, several factors could contribute to discrepancies between true and observed clinical incidence in the field. Gwembe HFCA, which otherwise has very low transmission, contains the district hospital. Clinical cases reported from Gwembe HFCA may have traveled from elsewhere within or outside the study area expressly to seek care. Case management rates in simulation were estimated from survey responses during the dry season, while individuals with fever during the wet season may show different health-seeking behavior based on road conditions and personal assessment of whether the fever is malarial. A strong distance-dependence on health-seeking (Figure S4B) means households closer to the health facility contribute disproportionately more to recorded case counts than to true clinical incidence. Lastly, clinics report any RDT-positive individual presenting with fever as a clinical case of malaria. However, a substantial portion of individuals in malaria-endemic areas with non-malarial fevers will be RDT-positive due to low-density infection or recent clearance of malaria.[20,21]

*Regional elimination requires concerted ITN distributions, drug campaigns, and improvement in case management*

Incomplete coverage and imperfect diagnostics in MSAT campaigns result in a significant portion of the parasite reservoir being left untreated.[21–25] This untreated reservoir will resume the cycle of transmission during the next rainy season. However, a mass-distribution mode that can treat undetected infections and a drug formulation such as DP with longer prophylactic protection against re-infection will be more successful at interrupting transmission.[10]

Operations teams in the Lake Kariba region continued mass treatment following the 2012-13 MSAT by administering MDA and focal MDA to a randomized group of HFCAs. Simulations approximated operational activities in 2014-15 by administering a mass distribution of ITNs in June 2014 and MDA with DP to all HFCAs in December 2014 and July 2015, in line with operational schedules.

To compare how case management, ITN usage, and MDA coverage contribute to reducing malaria burden separately and together, a variety of post-2015 intervention scenarios were simulated. For each simulation, the fraction of the study-area population living in clusters with prevalence below 1% was measured. Because simulated migration rates were high, this metric approximates the extent to which the entire study area has achieved elimination.

Discontinuing MDAs after the 2015 rounds while maintaining current levels of ITN usage did not result in long-term reduction in regional malaria transmission (Figure 4A). Increasing passive case management rates established a new baseline of lower transmission during the decade following the end of MDAs in 2015 but could not achieve region-wide elimination. Combining the 2014-15 MDAs with ramp-ups in case management and increased ITN usage resulted in near-elimination under highly aggressive ITN distribution campaigns but not if ITN ramp-ups followed historical rates of increase (Figure 4B, Figure S9). Under both ITN scenarios, transmission rebounded after the last ITN distribution in 2022 to a new baseline largely determined by case management rate. Coverage achieved during MDA did not affect transmission in the long term (Figure S10).

Extending annual dry-season MDAs through 2020 resulted in high probability of complete elimination if ITN usage was aggressively increased, but not if ITN usage was maintained at current levels or ramped up gradually (Figure 4C, Figure S11). Simulations predict that elimination in the Lake Kariba region requires both high ITN usage to reduce vectorial capacity and multi-year MDAs to deplete the parasite reservoir.

Restricting post-2015 MDAs to high-burden HFCAs resulted in outcomes comparable to cases where MDAs were distributed to all HFCAs (Figure 4D). Simulating with a lower migration rate under less aggressive ITN distributions showed a slower rebound in transmission after 2022, suggesting that success of targeted approaches is highly dependent on regional spatial connectivity.

Unsurprisingly, elimination was easier to achieve in clusters situated in low-burden HFCAs (Figure 4E). Several combinations of intervention mixes resulted in nearly all low-burden clusters maintaining very low prevalence through January 2028 and likely interrupting local transmission. Limiting migration reduced or eliminated transmission even in several high-burden clusters, suggesting that a few very high-burden clusters are maintaining transmission throughout the entire study area.

**Discussion**

The complex spatial dynamics of malaria transmission in the Lake Kariba region of Southern Zambia were captured in a mathematical model, which was then extended to compare the effects of future interventions. The simulation framework was informed by household surveillance in 2012-13, clinical case counts reported by local clinics, and preliminary entomological data. This work is the first to examine how village-scale variation in transmission intensity can drive transmission throughout an interconnected region.

While the 2012-13 MSAT surveys were a rich source of data for model-building, they presented an incomplete picture of local malaria transmission because measurements were taken during the dry season. The seasonality of transmission was thus difficult to characterize based solely on these measurements. Weekly clinical case counts reported from health facilities were broadly informative of transmission during the remaining months of the year, but ascertaining the proportion of RDT-positive fevers that are true malarial fevers is difficult. The seasonality of cases observed at the health facility may not reflect seasonality of transmission in every village within the HFCA as each HFCA may encompass a range of transmission intensities. Nonetheless, simulations were able to recapture both region- and village-scale spatio-temporal features of transmission observed during the 2012-13 operations.

Local entomology drives the seasonality of transmission. Attempting to model the study area with only *arabiensis* vectors requires non-seasonal year-round transmission in many lakeside areas in order to capture substantial prevalence during the November surveys, which would result in a mismatch with observed seasonality of clinical case counts. Incorporating *funestus* allowed the model to capture both dry season prevalence and seasonal patterns of clinical cases. More entomological data on vector species abundance and behavior will guide continued model refinements.

All simulations assumed full compliance with drug regimens. While perfect adherence to treatment is unlikely,[26,27] elimination is possible even when compliance is extremely poor (Figure S12), and previous modeling work has shown that compliance is not crucial in mass campaigns in settings without drug resistance.[10] Most individuals who harbor dry season infections have low parasite densities,[20,21] and a single dose of AL or DP is often sufficient to clear those infections.

Unlike the 2012-13 MSATs, the 2014-15 MDAs correspond to a dramatic move across the region toward elimination. However, this effect is short-term unless coupled with improvements in passive case detection and vector control. In this study, vector control has been modeled exclusively as ITN usage, but indoor residual spraying (IRS) would have similar effects if local vectors are susceptible. Combining IRS and ITNs may likely represent a more attainable scenario for aggressive vector control given that ITN usage has a large and unpredictable behavioral component. Simulations predict that elimination is highly likely if MDAs are continued for a few more years and vector control remains very aggressive, although as predicted by generic models,[28] even in this context transmission can occasionally rebound after vector control ceases (Figure S11). Rebound in transmission depends on the extent of human movement as travelers from high-transmission areas re-seed transmission in low-transmission areas.

Targeting MDAs to high-burden areas is potentially a reasonable approach to saving resources while achieving comparable outcomes to non-targeted approaches, and could even result in improved outcomes if better coverage with vector control can be achieved in targeted areas. Simulations predict that as long as passive case management improves, malaria transmission will die out in low-burden areas. While generic models have predicted that targeting transmission foci would be effective,[29] this study demonstrates for the first time that elimination strategies in interconnected regions are better served by focusing directly on reducing transmission in regional hotspots rather than attempting to maintain an "elimination front" and tackling the most difficult areas last. Programs will need to identify where regional hotspots are located, with the understanding that hotspot locations and intensities can change dynamically as interventions are deployed and changes in vector species composition, human movement patterns, housing conditions, and climatic cycles alter the local landscape of transmission.

As more regions reduce transmission and approach malaria elimination, it becomes crucial to understand how to set up malaria operations for successful and lasting elimination at local, national, and regional levels. Southern Africa is a particular challenge as vectorial capacity is substantial in some areas and the region has become increasingly interconnected. This study suggests that elimination in the southern Africa region will require several years of concerted vector control and mass drug campaigns that systematically target local transmission hotspots within the region. Ultimately, it follows that regional malaria elimination in southern Africa is nevertheless within reach with current tools, provided the efficacy and operational efficiency attained in the Lake Kariba operational area can be extended and targeted to other key areas.

**Author contributions**

PAE and EAW conceived the study. BH and JMM provided the data. CAB and AUB conducted the data analysis. MN, EAW, and JG designed the simulations. MN performed the simulations. MN, CAB, AUB, PAE, EAW, and JG prepared the manuscript.

**Acknowledgments**

We would like to acknowledge the support of the Zambia Ministry of Health, National Malaria Control Centre, the Provincial Health Office, and local district health offices in conducting the data collection. We are grateful to the communities involved in the research for their continued support in the quest to eliminate malaria. This work was funded by Bill and Melinda Gates through the Global Good Fund and the Bill and Melinda Gates Foundation (OPP 1089412 to PATH MACEPA).

**Declaration of interests**

All authors declare that they have no competing interests.

**References**

1	World Health Organization. World Malaria Report 2015. Geneva: World Health Organization, 2015.
2	Bhatt S, Weiss DJ, Cameron E, et al. The effect of malaria control on *Plasmodium falciparum* in Africa between 2000 and 2015. *Nature* 2015; **526**: 207–11.
3	Poirot E, Skarbinski J, Sinclair D, Kachur SP, Slutsker L, Hwang J. Mass drug administration for malaria. *Cochrane Database Syst Rev* 2013; **12**: CD008846.
4	Tatem AJ, Smith DL, Gething PW, Kabaria CW, Snow RW, Hay SI. Ranking of elimination feasibility between malaria-endemic countries. *Lancet* 2010; **376**: 1579–91.
5	Tatem AJ, Smith DL. International population movements and regional *Plasmodium falciparum* malaria elimination strategies. *Proc Natl Acad Sci U S A* 2010; **107**: 12222–7.
6	Cotter C, Sturrock HJ, Hsiang MS, et al. The changing epidemiology of malaria elimination: new strategies for new challenges. *Lancet* 2013; **382**: 900–11.
7	Larsen DA, Bennett A, Silumbe K, et al. Population-Wide Malaria Testing and Treatment with Rapid Diagnostic Tests and Artemether-Lumefantrine in Southern Zambia: A Community Randomized Step-Wedge Control Trial Design. *Am J Trop Med Hyg* 2015; **92**: 913–21.
8	Larsen DA, Chisha Z, Winters B, et al. Malaria surveillance in low-transmission areas of Zambia using reactive case detection. *Malar J* 2015; **14**: 465.
9	Griffin JT, Hollingsworth TD, Okell LC, et al. Reducing *Plasmodium falciparum* Malaria Transmission in Africa: A Model-Based Evaluation of Intervention Strategies. *PLoS Med* 2010; **7**: e1000324.
10	Gerardin J, Eckhoff P, Wenger EA. Mass campaigns with antimalarial drugs: a modelling comparison of artemether-lumefantrine and DHA-piperaquine with and without primaquine as tools for malaria control and elimination. *BMC Infect Dis* 2015; **15**: 144.
11	Maude RJ, Socheat D, Nguon C, et al. Optimising strategies for *Plasmodium falciparum* malaria elimination in Cambodia: primaquine, mass drug administration and artemisinin resistance. *PLoS One* 2012; **7**: e37166.
12	Gu W, Killeen GF, Mbogo CM, Regens JL, Githure JI, Beier JC. An individual-based model of *Plasmodium falciparum* malaria transmission on the coast of Kenya. *Trans R Soc Trop Med Hyg* 2003; **97**: 43–50.
13	Silal SP, Little F, Barnes KI, White LJ. Hitting a Moving Target: A Model for Malaria Elimination in the Presence of Population Movement. *PLoS One* 2015; **10**: e0144990.
14	Eckhoff PA. A malaria transmission-directed model of mosquito life cycle and ecology. *Malar J* 2011; **10**: 303.
15	Eckhoff P. *P. falciparum* Infection Durations and Infectiousness Are Shaped by Antigenic Variation and Innate and Adaptive Host Immunity in a Mathematical Model. *PLoS One* 2012; **7**: e44950.
16	Rainfall Estimator 2.0. http://www.cpc.ncep.noaa.gov/products/fews/data.shtml (accessed July 15, 2015).
17	Bhatt S, Weiss DJ, Mappin B, et al. Coverage and system efficiencies of insecticide-treated nets in Africa from 2000 to 2017. *eLife* 2015; **4**. DOI:10.7554/eLife.09672.
18	Chanda J. MACEPA vector bionomics in Southern Province of Zambia [online]. Email to Edward Wenger 2015 Sept 24.
19	Killeen GF, McKenzie FE, Foy BD, Bøgh C, Beier JC. The availability of potential hosts as a determinant of feeding behaviours and malaria transmission by African mosquito populations. *Trans R Soc Trop Med Hyg* 2001; **95**: 469–76.

20  Okell LC, Bousema T, Griffin JT, Ouédraogo AL, Ghani AC, Drakeley CJ. Factors determining the occurrence of submicroscopic malaria infections and their relevance for control. *Nature Comm* 2012; **3**: 1237.

21  Gerardin J, Ouédraogo AL, McCarthy KA, Eckhoff PA, Wenger EA. Characterization of the infectious reservoir of malaria with an agent-based model calibrated to age-stratified parasite densities and infectiousness. *Malar J* 2015; **14**: 231.

22  Tiono AB, Ouédraogo A, Ogutu B, et al. A controlled, parallel, cluster-randomized trial of community-wide screening and treatment of asymptomatic carriers of *Plasmodium falciparum* in Burkina Faso. *Malar J* 2013; **12**: 79.

23  Slater HC, Ross A, Ouédraogo AL, et al. Assessing the impact of next-generation rapid diagnostic tests on *Plasmodium falciparum* malaria elimination strategies. *Nature* 2015; **528**: S94–S101.

24  Halliday KE, Okello G, Turner EL, et al. Impact of Intermittent Screening and Treatment for Malaria among School Children in Kenya: A Cluster Randomised Trial. *PLoS Med* 2014; **11**: e1001594.

25  Kern SE, Tiono AB, Makanga M, et al. Community screening and treatment of a symptomatic carriers of *Plasmodium falciparum* with artemether-lumefantrine to reduce malaria disease burden: a modelling and simulation analysis. *Malar J* 2011; **10**: 210.

26  Yeung S, White NJ. How do patients use antimalarial drugs? A review of the evidence. *Trop Med Int Health* 2005; **10**: 121–38.

27  Minzi O, Maige S, Sasi P, Ngasala B. Adherence to artemether-lumefantrine drug combination: a rural community experience six years after change of malaria treatment policy in Tanzania. *Malar J* 2014; **13**: 267.

28  Briët OJ, Penny MA. Repeated mass distributions and continuous distribution of long-lasting insecticidal nets: modelling sustainability of health benefits from mosquito nets, depending on case management. *Malar J* 2013; **12**: 401.

29  Bousema T, Griffin JT, Sauerwein RW, et al. Hitting Hotspots: Spatial Targeting of Malaria for Control and Elimination. *PLoS Med* 2012; **9**: e1001165.

**Figure legends**

Figure 1. Households in the Lake Kariba region of Southern Province, Zambia, are clustered into village-scale simulation constructs within twelve health facility catchment areas.

Figure 2. Spatial distribution of two vector species is calibrated to prevalence survey and clinical case count data. (A) Adult vector numbers are tuned by scaling larval habitat availability of *An. arabiensis* and *An. funestus* vectors. Relative scales of the two vector species govern the seasonality of human biting. Shown: *arabiensis* scale = 100, *funestus* scale = 30. (B) Representative calibration of larval habitat availabilities in a Chabbobboma HFCA cluster. Sampling over larval habitat availability scale factors for *arabiensis* and *funestus* identifies combinations of scale factors with good fits to RDT prevalence from surveillance data. Incorporating scores from comparisons of cluster-level simulated clinical case counts to HFCA-level reported clinical case counts results in improved fit to the seasonality of clinical incidence while maintaining good fit to observed prevalence. Cluster-level simulated clinical case counts were scaled to match HFCA-wide populations. (C) Best fit *arabiensis* and *funestus* larval habitat availabilities

vary spatially over the study region, with more habitat available for both species in the lower-altitude regions closer to the lake front.

Figure 3. Simulated malaria transmission in the Lake Kariba region shows good agreement with surveillance data. (A) Spatio-temporal variation in cluster-level prevalence of RDT-positive infections is well-captured by simulation. Top: survey data collected during MSAT rounds. Bottom: mean cluster prevalence in ten stochastic simulation realizations. (B) Simulations capture the seasonality of HFCA-level weekly clinical case counts. Simulated clinical and severe malaria cases are scaled by 32% to estimate the fraction seeking treatment in clinics. Mean (line) and range (shaded area) of ten stochastic realizations.

Figure 4. Malaria elimination in the Lake Kariba region requires deployment of mass drug campaigns, high levels of ITN usage, and improved case management rates. (A) Transmission reduction if no MDAs are conducted after 2015 and ITN usage is not increased. Degree of regional elimination is measured as the fraction of the total population living in village-clusters with RDT prevalence below 1%. All traces show the mean (line) and range (shaded area) of ten stochastic realizations. See Supplemental Methods for details. (B) Transmission reduction if no MDAs are conducted after 2015, case management is ramped up, and ITN usage is increased between 2015-22. (C) Transmission reduction if MDAs continue through 2020 and case management is ramped up. (D) Transmission reduction if MDAs between 2016 and 2020 are distributed only to high-burden HFCAs and case management is ramped up. (E) Cluster RDT prevalence in December 2027 for ten possible scenarios of future interventions. Clusters are divided into high-burden (n=64) and low-burden (n=51) groups as indicated in Fig 4D. Cluster prevalence is the mean of ten stochastic realizations.

**Supporting Information**
Supplemental Methods
Figure S1. Surveillance and simulation demographics.
Figure S2. Spatial and seasonal variation in climate used in simulation.
Figure S3. ITN usage in surveillance and simulation.
Figure S4. Health-seeking in surveillance and simulation.
Figure S5. MSAT coverage by HFCA.
Figure S6. MSAT coverage can vary substantially within an HFCA.
Figure S7. Prevalence correction based on differences in demographics between surveillance data and simulation.
Figure S8. Surveillance data and simulated cluster RDT prevalence across six MSAT rounds.
Figure S9. ITN ramp-up trajectories used in post-2016 intervention scenarios.
Figure S10. Coverage achieved during 2014-15 MDAs has little effect on long-term transmission.
Figure S11. MDAs through 2020 coupled with aggressive ITN distributions result in high likelihood of elimination.
Figure S12. Elimination is less likely under extremely poor compliance with MDA treatment.

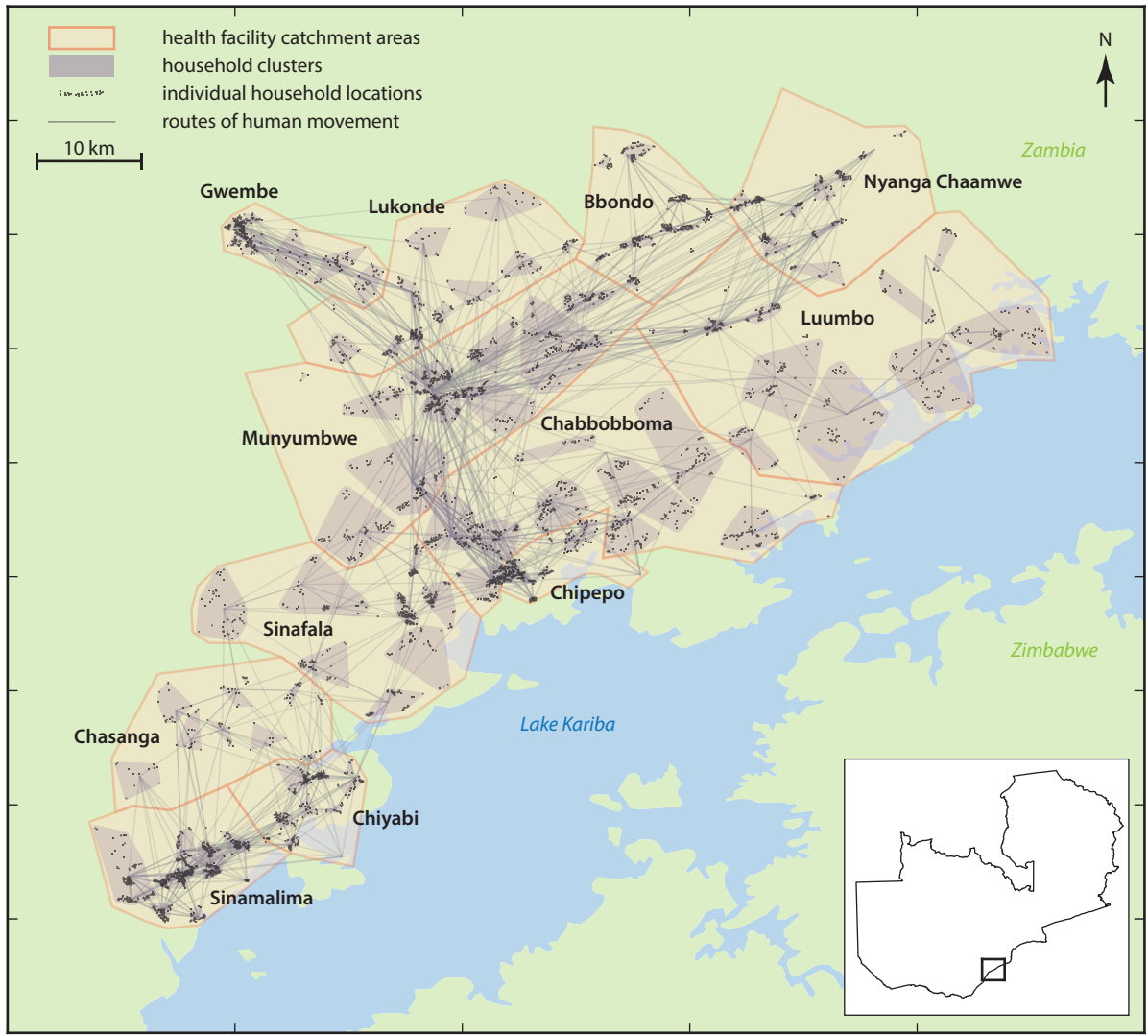

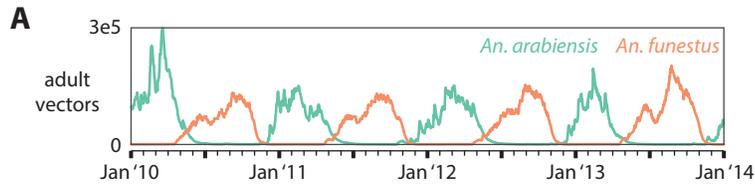
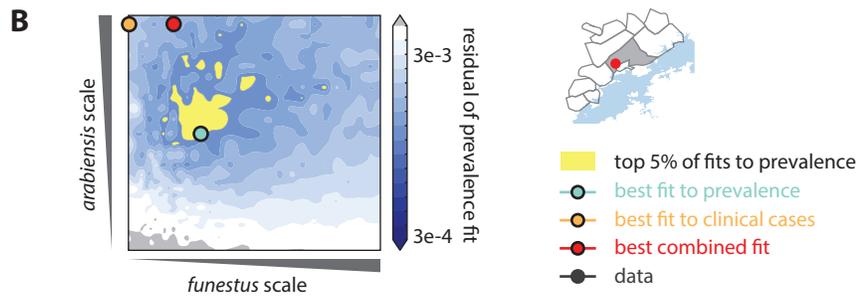
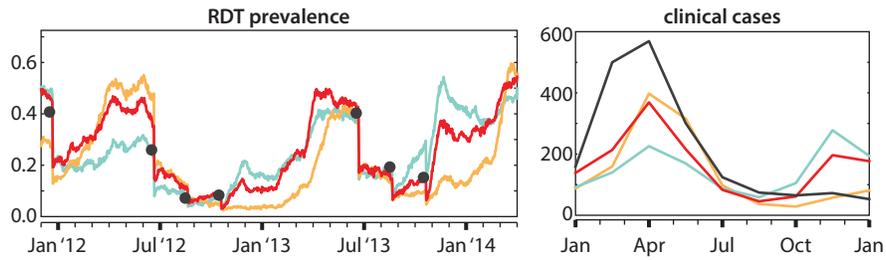
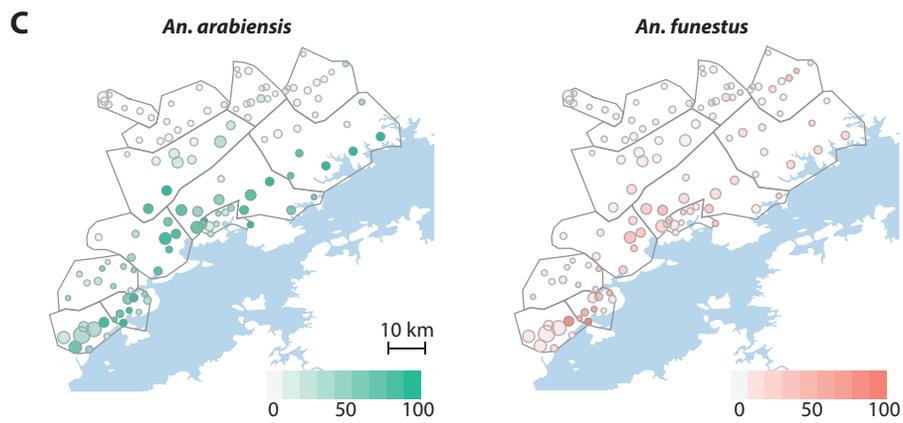

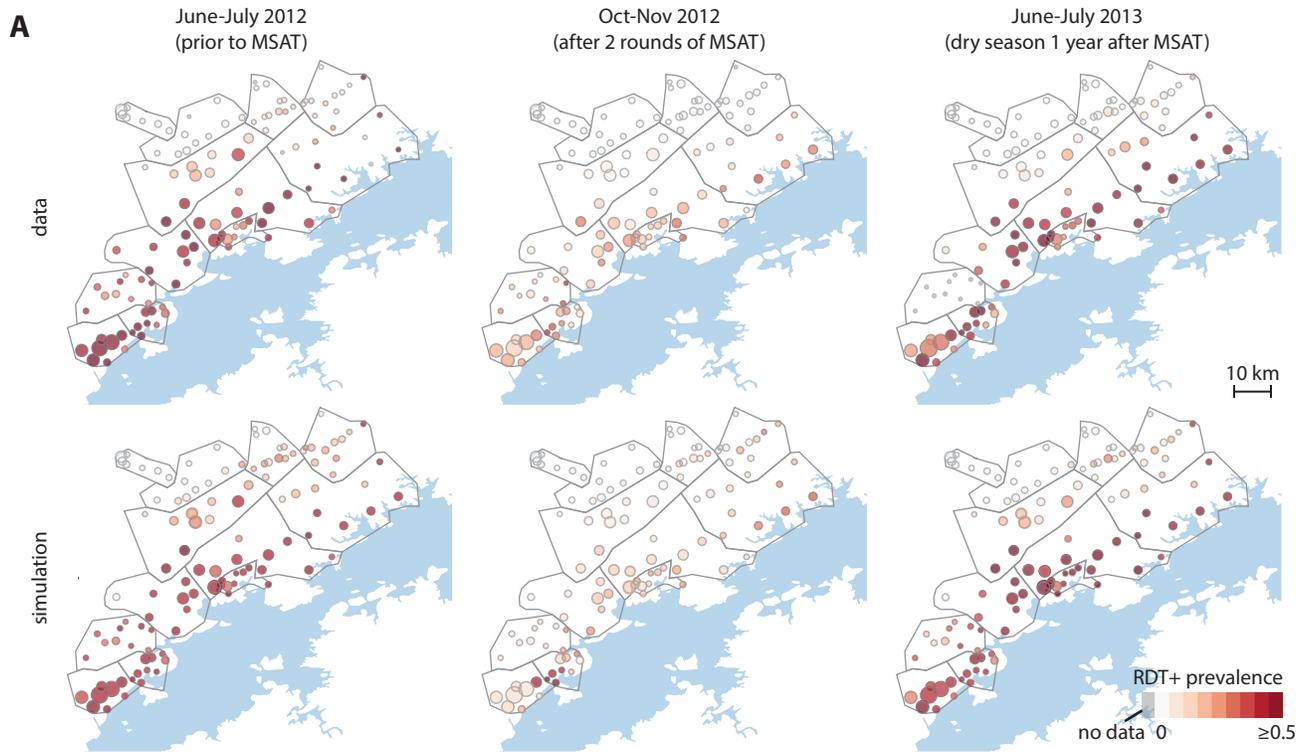
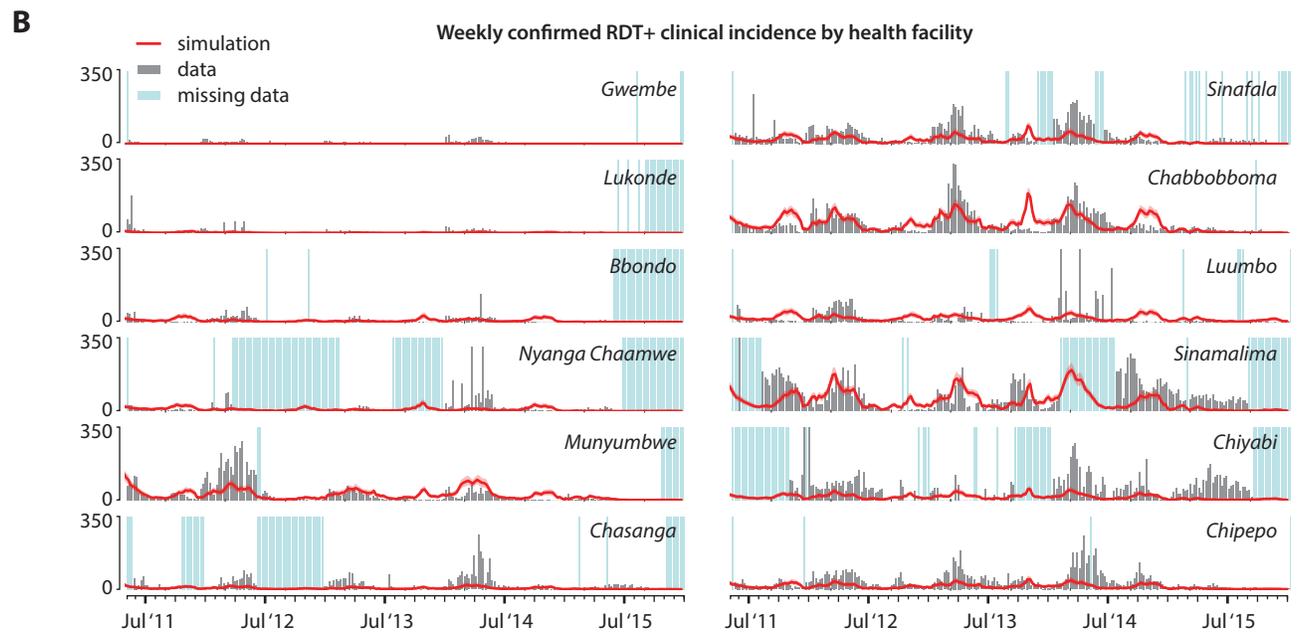

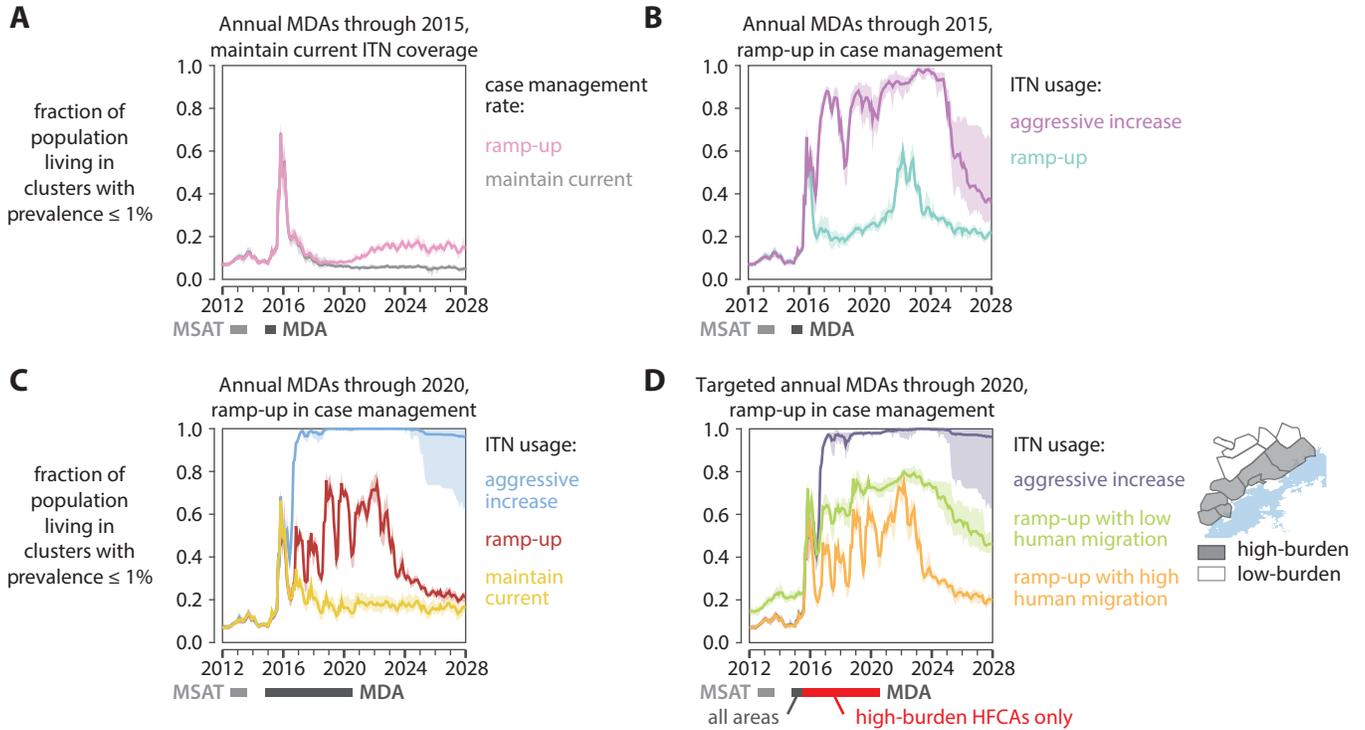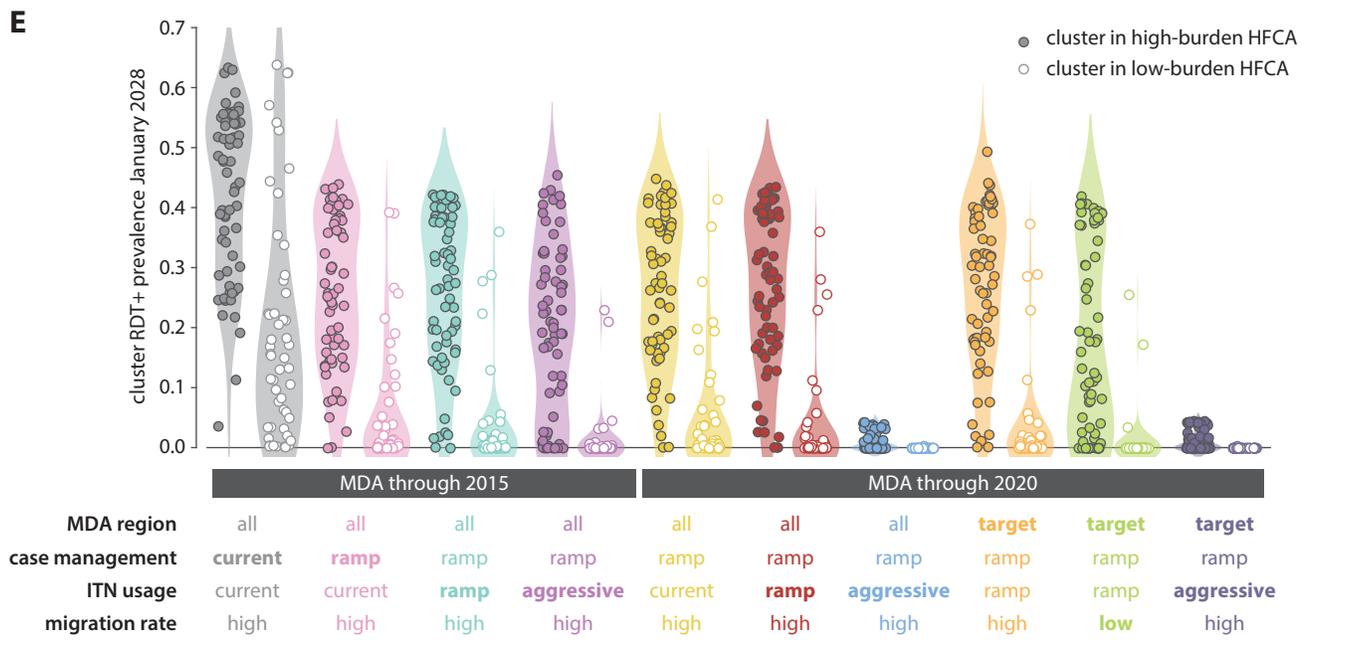